\documentclass{ws-procs975x65}            
\begin{document}
\title{Supernova Bounds on keV-mass Sterile Neutrinos}

\author{Shun Zhou$^*$}

\address{Institute of High Energy Physics, Chinese Academy of Sciences,
Beijing 100049, China\\
$^*$E-mail: zhoush@ihep.ac.cn}

\begin{abstract}
Sterile neutrinos of keV masses are one of the most promising candidates for the warm dark matter, which could solve the small-scale problems encountered in the scenario of cold dark matter. We present a detailed study of the production of such sterile neutrinos in a supernova core, and derive stringent bounds on the active-sterile neutrino mixing angles and sterile neutrino masses based on the standard energy-loss argument.
\end{abstract}


\bodymatter

\section{Motivation}

The observations of rotational curves of galaxies, the bullet cluster, gravitational lensing effects and cosmic microwave background have provided us with robust evidence that the matter content of our Universe
is dominated by the non-baryonic dark matter. So far, a lot of attention has been focused on the scenario of cold dark matter (CDM), which has a
negligible velocity dispersion at the radiation-matter equality and damps structures below the Earth-mass scales.\cite{CDM1} The favourite candidates for CDM stem from the well-motivated theories, which have been proposed to solve fundamental problems of the standard model of elementary particle physics,\cite{DMRev2} such as the lightest supersymmetric particle and the
axion. However, the CDM scenario suffers from a few serious problems in the galaxy and small-scale structure formation, e.g., the overprediction of the observed satellites in the galaxy-scale halos\cite{CDM2} and the high
concentration of dark matter in galaxies.\cite{CDM3} In the scenario
of warm dark matter (WDM), a light-mass particle with a large
velocity dispersion can suppress the structure formation up to the
galaxy scales and thus solve the potential small-scale structure problems.\cite{Colombi:1995ze}

Sterile neutrinos of keV masses are a promising candidate for the
WDM. \cite{SteRev1} As a simple but instructive example, Dodelson and Widrow have proposed that right-handed neutrinos with masses $m_s \sim {\rm keV}$ can be produced via neutrino oscillations in the early Universe and account for all the dark matter,\cite{DW} if they mix with the ordinary neutrinos via a tiny mixing angle $\theta \sim 10^{-(4\cdots 6)}$ in vacuum. Due to such a small mixing angle, sterile neutrinos can never be in thermal
equilibrium. In the presence of a primordial lepton asymmetry, Shi and
Fuller have observed that the production rate of sterile neutrinos
could be enhanced by the Mikheyev-Smirnov-Wolfenstein (MSW) effect
\cite{MSW1,MSW2} and the correct relic abundance can be obtained even for
much smaller mixing angles.\cite{SF} One possible way to detect
sterile-neutrino WDM is to search for the X-rays from their
radiative decays in the DM-dominated galaxies. \cite{Abazajian} Conversely, the non-observation of an X-ray line from the local group dwarf galaxies has placed restrictive limits on the mass and mixing angle of sterile
neutrinos. Recently, the observation of an X-ray line around $3.5~{\rm keV}$ in the Andromeda galaxy and the Perseus galaxy cluster has been claimed by two independent groups.\cite{Bulbul:2014sua,Boyarsky:2014jta} However, this claim is criticized by other authors, who attribute this X-ray line to the Potassium and Chlorine emission lines from the plasma in the galactic regions and the intergalactic gas.\cite{Jeltema:2014qfa} At the present time, it is better to leave this problem open, and the future X-ray data could hopefully offer us a clue. Other limits can be obtained from the observations of Lyman-alpha forest and Supernova (SN) 1987A.\cite{SteRev1} Put all together, the window for the Dodelson-Widrow mechanism of non-resonant production is closed, while the Shi-Fuller mechanism of resonant production is still viable.\cite{constraint} Roughly speaking, sterile neutrinos with $m_s = {1\sim  10}~{\rm keV}$ and $\theta = 10^{-(4\cdots 6)}$ could be WDM. However, it should be noticed that the observational constraints depend crucially on the production mechanisms and evolution of sterile neutrinos in the early Universe, so the constraints can be evaded as in several new-physics models.\cite{model1,model2,Merle:2013gea}

Moreover, the WDM sterile neutrinos could play an important role in
generating the supernova asymmetries for pulsar kicks,\cite{pulsar} and in supporting supernova explosions.\cite{Hidaka1,Hidaka2} Hence it is interesting to reexamine the SN bounds on the keV-mass sterile neutrinos by studying in detail the production and propagation of sterile neutrinos in the SN core. On the other hand, as we show later, the phenomenon of neutrino oscillations and interactions in a dense medium is intriguing by itself.

\section{General Formalism}

The matter density in a SN core $\rho = 3.0\times 10^{14}~{\rm g}~{\rm cm}^{-3}$ is so high that even the weakly-interacting neutrinos could not escape freely.\cite{Bethe:1990mw} Hence the production of sterile neutrinos in the dense matter can be very efficient through both neutrino oscillations and repeated scattering of ordinary neutrinos off background particles. In consideration of both coherent flavor oscillations and the decoherence caused by frequent scattering, it is convenient to describe the whole neutrino system in terms of density matrices.\cite{formalism,formalism1} We assume that neutrino interactions do not affect substantially the medium, whose configuration is mainly determined by the conditions of thermal equilibrium. The time duration of collisions between neutrinos and matter particles is much shorter than the evolution time scale, on which neutrino density matrices experience a significant change. On the other hand, the evolution of density matrices is sufficiently rapid compared to the macroscopic and hydrodynamic time scale.

Under the above assumptions, we follow the formalism mainly developed by Raffelt and Sigl,\cite{formalism,formalism1} and define the ensemble average of the density matrices of $n$-flavor neutrinos as $\langle a^\dagger_j({\bf p}) a^{}_i({\bf q})\rangle = (2\pi)^3 \delta^3({\bf p} - {\bf q}) (\rho_{\bf p})^{}_{ij}$ and $\langle b^\dagger_i({\bf p}) b^{}_j({\bf q})\rangle = (2\pi)^3 \delta^3({\bf p} - {\bf q}) (\bar{\rho}_{\bf p})^{}_{ij}$ for $i, j = 1, 2, \cdots, n$, where $a^{}_i({\bf p})$ and $b^{}_i({\bf p})$ stand for the annihilation operators for neutrinos and antineutrinos, respectively. The diagonal elements of $\rho^{}_{\bf p}$ and $\bar{\rho}^{}_{\bf p}$ are just the occupation numbers, while the off-diagonal elements encode the phase information. The evolution of the matrix of densities $\rho^{}_{\bf p}$ is governed by\cite{formalism,formalism1}
\begin{eqnarray}
\dot{\rho}^{}_{\bf p} = -{\rm i} \left[\Omega^{}_{\bf p}, \rho^{}_{\bf p}\right] &+& \frac{1}{2} \sum^n_{i = 1} \left[ \{I^{}_i, 1 - \rho^{}_{\bf p}\} {\cal P}^i_{\bf p} - \{I^{}_i, \rho^{}_{\bf p}\} {\cal A}^i_{\bf p} \right] \nonumber \\
&+& \frac{1}{2} \sum_a \int \frac{{\rm d}^3{\bf p}^\prime}{(2\pi)^3} \left\{ \left[G^a \rho^{}_{{\bf p}^\prime} G^a (1 - \rho^{}_{\bf p}) + {\rm h.c.}\right] {\cal W}^a_{{\bf p}^\prime {\bf p}} \right. \nonumber \\
&~& ~~~~~~~~~~~~~~~~~ - \left. \left[\rho^{}_{\bf p} G^a (1 - \rho^{}_{{\bf p}^\prime}) G^a  + {\rm h.c.}\right] {\cal W}^a_{{\bf p} {\bf p}^\prime} \right\} \; .
\end{eqnarray}
On the right-hand side, the first term contains neutrino kinetic energy and matter potential from the coherent forward scattering. The second term arises from the charged-current production and absorption of neutrinos, where $I^{}_i$ is the projection matrix onto the neutrino flavor $i$, ${\cal P}^i_{\bf p}$ the production rate, and ${\cal A}^i_{\bf p}$ the absorption rate. The third term represents the contribution from scattering processes $\nu^{}_{\bf p} + a \to \nu^{}_{{\bf p}^\prime} + a^\prime$ with $a$ and $a^\prime$ being any kind of background particles, where $G^a$ is a diagonal $n\times n$ matrix of couplings $g^a_i$ for the interaction type $a$ and $(g^a_i)^2 {\cal W}^a_{{\bf p}^\prime {\bf p}}$ denotes the transition probability. In the case of antineutrinos, the evolution equation for $\bar{\rho}^{}_{\bf p}$ can be obtained in a similar way.

Applying this general formalism to an ensemble of a keV-mass sterile neutrino mixing with only one flavor of ordinary neutrinos, we can simplify significantly the Boltzmann-type equations. To further simplify the picture, we consider a homogeneous and isotropic SN core, in which neutrinos are essentially trapped and well in thermal equilibrium with the medium. Due to the isotropy of the medium and neutrino ensemble, the relevant interaction rates depend only on neutrino energies. The next important step is to average $(\rho^{}_E)^{}_{ij}$ and $(\bar{\rho}^{}_E)^{}_{ij}$ over many cycles of flavor oscillations, since the oscillation length is much shorter than neutrino mean free path. More explicitly, we can estimate the oscillation length as
\begin{equation}
\lambda^{}_{\rm osc} \lesssim 0.7~{\rm cm} \left(\frac{E}{30~{\rm MeV}}\right) \left(\frac{10^{-4}}{\sin 2\theta}\right) \left(\frac{10~{\rm keV}}{m^{}_s}\right)^2 \; ;
\end{equation}
and the mean-free-path as
\begin{equation}
\lambda^{}_{\rm mfp} = \frac{1}{N^{}_{\rm B} \sigma^{}_{\nu N}} \approx 1.1\times 10^3~{\rm cm} \left(\frac{30~{\rm MeV}}{E}\right)^2 \rho^{-1}_{14} \; ,
\end{equation}
where $N^{}_{\rm B}$ denotes the number of baryons, $\sigma^{}_{\nu N} \sim G^2_{\rm F} E^2/\pi$ the cross section of neutrino-nucleon interaction and $\rho^{}_{14}$ the matter density in units of $10^{14}~{\rm g}~{\rm cm}^{-3}$. For a typical mixing angle $\sin^2 2\theta = 10^{-8}$ and sterile neutrino mass $m^{}_s = 10~{\rm keV}$, we always have $\lambda^{}_{\rm mfp} \gg \lambda^{}_{\rm osc}$, corresponding to the weak-damping limit. This comparison indicates that neutrinos oscillate a large number of times before a subsequent scattering with nucleons. The averaged matrix of densities $\tilde{\rho}^{}_E$ can now be parameterized in terms of the distribution functions of active and sterile neutrinos, denoted as $f^\alpha_E$ and $f^s_E$, respectively. If the active neutrino stays in thermal equilibrium and if the mixing angle is so small that sterile neutrinos escape from the SN core immediately after production, the evolution equation of $f^s_E$ is given by\cite{formalism1}
\begin{equation}
\dot{f}^s_E = \frac{1}{4} s^2_{2\theta_\nu} \left[{\cal P}^{}_E + \int \frac{{E^\prime}^2 {\rm d}E^\prime}{2\pi^2} g^2_\alpha {\cal W}^{}_{E^\prime E} f^\alpha_{E^\prime} \right] \; ,
\end{equation}
where $s^{}_{2\theta_\nu} \equiv \sin 2\theta^{}_\nu$ with $\theta_\nu$ being the neutrino mixing angle in matter, $g^{}_\alpha$ is the coupling constant for active neutrinos. Therefore, the emission rate of lepton number can be obtained by integrating the above equation over sterile neutrino energy. Similarly, the energy-loss rate can be calculated by convolving Eq. (4) with neutrino energy and integrating them over the whole energy range.

\section{Sterile neutrinos in SN cores}

Sterile neutrinos with masses in the keV range can be copiously produced
in the SN core. For $m_s \gtrsim 100~{\rm keV}$, the vacuum mixing angle of sterile neutrinos is stringently constrained $\sin^2 2\theta \lesssim 10^{-9}$ in order to avoid excessive energy loss.\cite{loss1,loss2,loss3} For smaller masses, however, the MSW effect on active-sterile neutrino mixing becomes very important and the SN bound on vacuum mixing angle is not that obvious. Note that the bounds on mixing angles depend on which neutrino species the sterile neutrino mixes with. In the following, we shall discuss $\nu^{}_\tau$-$\nu_s$ and $\nu_e$-$\nu_s$ mixing cases, and emphasize the main difference between them.

\subsection{$\nu_\tau$-$\nu_s$ mixing}

First, we concentrate on the SN bound in the simplest case of $\nu_\tau$-$\nu_s$ mixing, because $\nu_\tau$ and $\overline{\nu}_\tau$ only have neutral-current interactions and essentially stay in thermal
equilibrium with the ambient matter. In the weak-damping limit, which is always valid for supernova neutrinos mixing with keV-mass sterile neutrinos, the evolution of $\nu_\tau$ number density is determined by\cite{RZ}
\begin{equation}
\dot{N}_{\nu_\tau} = -\frac{1}{4} \sum_a \int \frac{E^2 {\rm
    d}E}{2\pi^2} s^2_{2\theta_\nu} \int \frac{{E^\prime}^2 {\rm
    d}E^\prime}{2\pi^2} W^a_{E^\prime E} f^\tau_{E^\prime} \; ,~~
\end{equation}
where $f^\tau_E$ is the occupation number of $\nu_\tau$, and ${\cal W}^a_{E^\prime E}$ the transition probability for $\nu(E^\prime) + a \to \nu(E) + a$ with $a$ being background particles in the SN core. Note that Eq.~(5) is actually an immediate consequence of integrating Eq.~(4) for $\nu_\tau$-$\nu_s$ mixing over the sterile neutrino energy. For $\nu^{}_\tau$, the coupling constant is just $g_\tau = 1$, so we write the transition probability as $W^a_{E^\prime E} = {\cal W}^a_{E^\prime E}$. The interaction types include neutrino-nucleon and neutrino-electron scattering, whereas the former dominates over the latter. In a similar way, we can derive the evolution equation of the $\bar\nu_\tau$ number density,
involving the mixing angle $\theta_{\bar\nu}$, the occupation number
$f^{\bar\tau}_E$ and the transition probability $\bar{W}^a_{E^\prime
E}$.

Due to the MSW effect, the mixing angle of neutrinos in matter
is different from that of antineutrinos, i.e.,
\begin{equation}
\sin^2 2\theta_{\nu,\bar\nu} = \frac{\sin^2 2\theta}{\sin^2 2\theta
+ (\cos
  2\theta \mp E/E_{\rm r})^2} \;,
\end{equation}
where $\theta$ denotes the vacuum mixing angle, and the sign ``$\mp$"
refers to $\nu$ and $\bar\nu$. However, the actual sign depends on the lepton asymmetries in the matter, i.e., the matter potential $V^{}_{\nu_\tau}$. The resonant energy $E_{\rm r} \equiv \Delta m^2/(2|V_{\nu_\tau}|)$ can be written as
\begin{equation}
E_{\rm r} = 3.25~{\rm MeV} \left(\frac{m_s}{10~{\rm
    keV}}\right)^2 \rho^{-1}_{14} \left|Y_0 - Y_{\nu_\tau}\right|^{-1} \; ,
\end{equation}
where $Y_0 \equiv (1 - Y_{e} - 2Y_{\nu_e})/4$, and $Y_x \equiv (N_x - N_{\bar x})/N_{\rm B}$
with $N_{\rm B}$ being the baryon number density, $N_x$ and
$N_{\bar{x}}$ being the number densities of particle $x$ and its
antiparticle $\bar x$. As for tau neutrinos, the matter potential
$V_{\nu_\tau} = - (G_{\rm F}/\sqrt{2}) N_{\rm B} \left(1 - Y_e -
2Y_{\nu_e} - 4Y_{\nu_\tau}\right)$ is negative if the typical values
of $Y_e = 0.3$, $Y_{\nu_e} = 0.07$ and $Y_{\nu_\tau} = 0$ for a SN
core are taken. Therefore, the mixing angle for $\bar\nu_\tau$ is
enhanced by matter effects, and the emission rate for $\bar\nu_\tau$
exceeds that for $\nu_\tau$, indicating that a
$\nu_\tau$-$\bar\nu_\tau$ asymmetry (i.e., $Y_{\nu_\tau} \neq 0$)
will be established. An interesting feedback effect emerges: (i) The
chemical potential for tau neutrinos develops and thus changes the
occupation numbers of $\nu_\tau$ and $\bar\nu_\tau$; (ii) The
$\nu_\tau$-$\bar\nu_\tau$ asymmetry shifts the resonant energy
$E_{\rm r}$, and thus modifies the mixing angles $\theta_\nu$ and
$\theta_{\bar\nu}$; (iii) Both effects in (i) and (ii) will feed
back on the emission rates. Hence a stationary state of this
active-sterile neutrino system could be achieved if the emission
rates for neutrinos and antineutrinos become equal to each other.\cite{RZ} However, whether such a stationary state can be really reached depends on the mixing angle and sterile neutrino mass. In any case, it is straightforward to calculate the energy-loss rate by following the time evolution of the neutrino system.

\subsection{$\nu_e$-$\nu_s$ mixing}

Since there is a great similarity between muon and tauon neutrino interactions in a SN core, we expect that $\nu_\mu$-$\nu_s$ mixing case shares the same features as the previous case. Then, we turn to the $\nu_e$-$\nu_s$ mixing case, where the physical processes for the production and evolution of sterile neutrinos are quite different.

First, the charged-current interaction $e^- + p \rightleftharpoons \nu_e + n$ leads to a direct production of $\nu_e$, and an absorption of $\nu_e$ by the medium as well. In addition to the charged-current interaction, neutrino scatterings off degenerate electrons and non-degenerate nucleons are also important. Therefore, both terms on the right-hand side of Eq.~(4) exist for the case of $\nu_e$-$\nu_s$ mixing.

Second, the emission of $\nu_s$ reduces the electron lepton number in the SN core, which is in thermal and beta equilibrium. From the latter condition, we can get a relationship $\mu_e = \mu_{\nu_e} + \hat{\mu}$ among the electron chemical potential $\mu_e$, the electron-neutrino chemical potential $\mu_{\nu_e}$ and the difference $\hat{\mu} \equiv \mu_n - \mu_p$ between the neutron $\mu_n$ and proton $\mu_p$ chemical potentials. The parameter $\hat{\mu} = 50 \sim 100~{\rm MeV}$ is determinable from the equation of state of the nuclear matter. In our discussions, we assume the beta equilibrium is always guaranteed and the equation of state is not significantly modified by the emission of sterile neutrinos. Under these assumptions, $\hat{\mu} = 55~{\rm MeV}$ is taken to be valid throughout. The emission of sterile neutrinos affects the number density of $\nu_e$, and thus $\mu_{\nu_e}$ and $\mu_e$, which in turn modify the matter potential for neutrinos and the mixing angle in matter. The modification of neutrino mixing angle obviously feeds back on the emission rate.

Third, since $\nu_\tau$ and $\bar{\nu}_\tau$ are always generated in pairs, there is no asymmetry between them, namely $Y^{}_{\nu_\tau} = 0$, and likewise for $\nu_\mu$ and $\bar{\nu}_\mu$. As a result, the resonant energy $E^{}_{\rm r} = \Delta m^2/(2|V^{}_{\nu_e}|)$ for neutrino oscillations is found to be
\begin{equation}
E_{\rm r} = 13~{\rm MeV} \left(\frac{m_s}{10~{\rm
    keV}}\right)^2 \rho^{-1}_{14} \left|1 - 3Y^{}_e - 4 Y_{\nu_e}\right|^{-1} \; ,
\end{equation}
where $Y^{}_{\nu_\mu} = Y^{}_{\nu_\tau} = 0$ is implemented. For the initial conditions $\mu_e = 218~{\rm MeV}$ and $\mu_{\nu_e} = 163~{\rm MeV}$, one obtains $Y^{}_e = 0.3$ and $Y^{}_{\nu_e} = 0.07$ for the temperature $T = 30~{\rm MeV}$ and matter density $\rho = 3.0\times 10^{14}~{\rm g}~{\rm cm}^{-3}$. Unlike in the $\nu_\tau$-$\nu_s$ mixing case, the matter potential $V^{}_{\nu_e}$ is positive, implying the resonance is initially in the neutrino sector. Therefore, the emission rate of $\nu_s$ is at the beginning much larger than that of $\bar{\nu}_s$, and the lepton number decreases rapidly. But even small changes of $Y^{}_e$ and $Y^{}_{\nu_e}$ are able to drive $E^{}_{\rm r}$ to infinity, where $1 - 3 Y^{}_e - 4 Y^{}_{\nu_e} = 0$ is reached.

It is worthwhile to mention that a large chemical potential $\mu_{\nu_e}$ indicates the population of $\bar{\nu}_e$ is rare. The initial emission of $\nu_s$ and the corresponding energy loss in this period become crucial for us to draw any SN limits based on the energy-loss argument. When $E^{}_{\rm r}$ is approaching infinity, the mixing angles in matter for neutrinos and antineutrinos are equal, and both of them are close to the vacuum mixing angle. During the evolution, the matter potential flips its sign and the resonance moves to the antineutrino sector. However, the $\bar{\nu}_e$ number density is largely suppressed due to the large chemical potential, even the enhancement of antineutrino mixing angle cannot enlarge the total emission rate remarkably.

\section{SN bound on sterile neutrinos}
\begin{figure}[t]
\begin{center}
\includegraphics[scale=0.8]{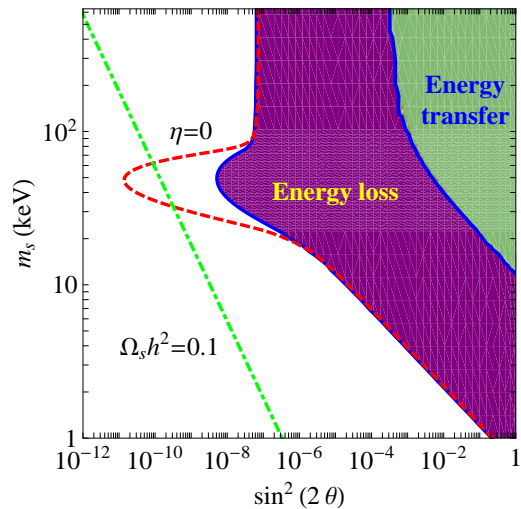}
\vspace{-0.1cm} \caption{Supernova bound on sterile neutrino masses
$m_s$ and active-sterile mixing angles $\theta$ in the $\nu_\tau$-$\nu_s$ mixing case, where the shaded purple region is excluded by the energy-loss argument while the shaded green one by the energy-transfer argument \cite{RZ}. The excluded region will be extended to the dashed (red) line if the build-up of degeneracy parameter is ignored, i.e., $\eta(t) = 0$. The dot-dashed (green) line represents the keV-mass sterile neutrinos as dark matter with the correct relic
abundance $\Omega_s h^2 = 0.1$. }
\end{center}
\end{figure}
Given the sterile neutrino mass $m_s$ and vacuum mixing angle
$\theta$, the energy loss rate ${\cal E}(t)$ due to sterile neutrino
emission can be calculated by following the evolution of
$\nu_\tau$-$\bar\nu_\tau$ asymmetry $Y_{\nu_\tau}(t)$. It has been
found that the stationary state can be reached within one second and
the feedback effect is very important for $20~{\rm keV} \lesssim m_s
\lesssim 80~{\rm keV}$ and $10^{-9} \lesssim \sin^2 2\theta \lesssim
10^{-4}$. To avoid excessive energy losses, we require that the
average energy-loss rate $\langle {\cal E} \rangle \equiv
\int^{\tau_{\rm d}}_0 {\cal E}(t)~{\rm d}t$ with $\tau_{\rm d} =
1~{\rm s}$ should be $\langle {\cal E} \rangle \lesssim 3.0\times
10^{33}~{\rm erg}~{\rm cm}^{-3}~{\rm s}^{-1}$. Otherwise, the
duration of neutrino burst from SN 1987A would have been
significantly reduced.

In Fig. 1, we show the contours of energy-loss rates in the $(\sin^2 2\theta, m_s)$-plane, where we have assumed a homogeneous and isotropic core with matter density $\rho = 3.0\times 10^{14}~{\rm g}~{\rm cm}^{-3}$ and temperature $T = 30~{\rm MeV}$. Based on the energy-loss argument, the shaded purple region has been excluded. The most stringent bound $\sin^2 2\theta \lesssim 10^{-8}$ arises for $m_s = 50~{\rm keV}$. For the
large-mixing angle region, the energy-loss rate is actually small,
because sterile neutrinos have been trapped in the core and cannot
carry energies away. However, the mean free path of sterile
neutrinos is comparable to or even larger than that of ordinary
neutrinos, indicating that they may transfer energies in a more
efficient way. As a consequence, the duration of neutrino burst will
be shortened by emitting neutrinos more rapidly. In this sense, the
excessive energy transfer should be as dangerous as the excessive
energy loss. Therefore, the large-mixing angle region is excluded
when the energy-transfer argument is applied. The green dot-dashed line in Fig. 1 indicates the relic abundance of dark matter $\Omega_s h^2 = 0.1$,
where keV-mass sterile neutrinos are warm dark matter and the
non-resonant production mechanism is assumed. If we ignore the
feedback effect (i.e., a vanishing chemical potential for tau
neutrinos $\eta = \mu_{\nu_\tau}/T = 0$), the excluded region will
extend to the red dashed line, which overlaps the relic-abundance
line. However, the mixing angles are essentially unconstrained in the
favored warm-dark-matter mass range $1~{\rm keV} \lesssim m_s \lesssim
10~{\rm keV}$.

\begin{figure}[t]
\begin{center}
\includegraphics[scale=1.0]{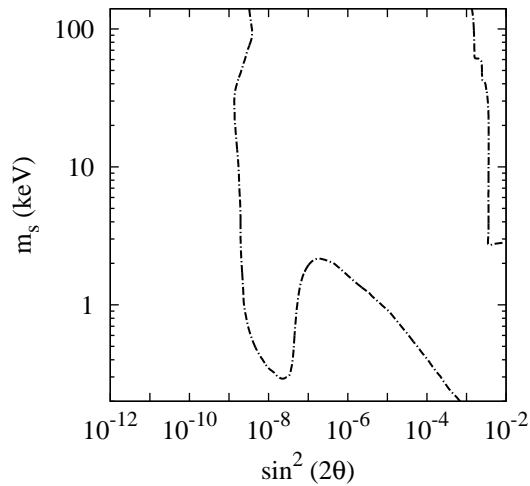}
\vspace{-0.1cm} \caption{Supernova bound on sterile neutrino masses
$m_s$ and active-sterile mixing angles $\theta$ in the $\nu_e$-$\nu_s$ mixing case. Based on the energy-loss argument, the parameter space between two dot-dashed curves is excluded. The region of large mixing angles at the upper-right corner is also disfavored if the energy-transfer argument is implemented.}
\end{center}
\end{figure}
As for the $\nu_e$-$\nu_s$-mixing case, we have included all the scattering processes of neutrinos, and solved the Boltzmann-like equation of the occupation numbers. In assumption of thermal equilibrium of $\nu_e$, one can derive the evolution equation of the degeneracy parameter $\eta = \mu_{\nu_e}/T$. It is straightforward to verify that the infinite $E^{}_{\rm r}$ is reached quickly, and the crossing from a negative to a positive matter potential takes place. In Fig. 2, using the same method as before, we have presented the contour plot of the energy-loss rate in the parameter plane of sterile neutrino masses and active-sterile neutrino mixing angles.\cite{CZ} On the upper-right corner, the energy-transfer argument can be applied to exclude this large-mixing angle region. For sterile neutrinos of masses above $1~{\rm keV}$, the leftmost exclusion line is nearly vertical around $\sin^2 2\theta = 10^{-9}$. For much smaller masses $m_s \ll 1~{\rm keV}$, the emission rate is highly suppressed by the matter effect, so the bound is rather loose.

\section{Summary and Outlook}

Since sterile neutrinos of keV masses as a promising candidate for warm dark matter are interesting in astrophysics and cosmology, we have considered their production in a SN core, and derived restrictive bounds on their masses and mixing angles by requiring the excessive energy loss caused by sterile neutrinos to be under control.

The main idea is to follow the time evolution of occupation numbers of neutrinos, and compute the emission rates of lepton number and sterile neutrino energy. Our final results about SN bounds are presented in Fig. 1 and Fig. 2. We have also emphasized the feedback effects in both $\nu_\tau$-$\nu_s$ and $\nu_e$-$\nu_s$ mixing cases: The emission of sterile neutrinos leads to changes in the lepton number asymmetries and thus the neutrino matter potentials, which in turn affect the emission rates via the MSW effects on neutrino mixing angles. In the case of $\nu_\tau$-$\nu_s$ mixing, the MSW resonance occurs in the antineutrino sector, and thus an asymmetry between $\nu_\tau$ and $\bar{\nu}_\tau$ number densities will be developed. This asymmetry contributes to the matter potential, intending to modify the mixing angles in matter and equating the emission rates of $\nu_s$ and $\bar{\nu}_s$. In the case of $\nu_e$-$\nu_s$ mixing, the matter potential $V^{}_{\nu_e}$ is rapidly driven to zero, implying that the neutrino and antineutrino mixing angles are close to the one in vacuum, which is in agreement with the result found in the literature.\cite{loss3} However, for much smaller masses, the emission rates will be suppressed by the matter effects, and the SN bound becomes very weak.

It is worth pointing out that keV-mass sterile neutrinos could also be produced in the collapsing phase of massive stars.\cite{Hidaka1,Hidaka2} Such sterile neutrinos after production propagate from the inner core to the outer one, and oscillate into electron neutrinos, which could deposit their energies in the out-layer matter and support SN explosions. A dedicated study of keV-mass sterile neutrinos in both core-collapse and cooling phases is certainly necessary and intriguing, and left for future works.



\begin{thebibliography}{99}
\bibitem{CDM1} S. Hofmann, D. J. Schwarz and H. Stoecker, Phys. Rev. D
  {\bf 64}, 083507 (2001).

\bibitem{DMRev2} F. D. Steffen, Eur. Phys. J. C. {\bf 59}, 557 (2009).

\bibitem{CDM2} G. Kauffmann, S. D. M. White and B. Guiderdoni,
  Mon. Not. R. Astron. Soc. {\bf 264}, 201 (1993).

\bibitem{CDM3} P. J. E. Peebles, arXiv:astro-ph/0101127.

\bibitem{Colombi:1995ze}
  S.~Colombi, S.~Dodelson and L.~M.~Widrow,
  Astrophys.\ J.\  {\bf 458}, 1 (1996)  [astro-ph/9505029].

\bibitem{SteRev1} A. Kusenko, Phys. Rept. {\bf 481}, 1 (2009).

\bibitem{DW} S. Dodelson and L. M. Widrow, Phys. Rev. Lett. {\bf 72},
  17 (1994).

\bibitem{MSW1} L. Wolfenstein, Phys. Rev. D {\bf 17}, 2369 (1978).

\bibitem{MSW2} S. P. Mikheyev and A. Yu. Smirnov, Sov. J. Nucl. Phys. {\bf 42}, 913 (1985).

\bibitem{SF} X. D. Shi and G. M. Fuller, Phys. Rev. Lett. {\bf 82}, 2832
  (1999).

\bibitem{Abazajian} K. N. Abazajian, arXiv:0903.2040 [astro-ph.CO].

\bibitem{Bulbul:2014sua}
  E.~Bulbul, M.~Markevitch, A.~Foster, R.~K.~Smith, M.~Loewenstein and S.~W.~Randall,
  Astrophys.\ J.\  {\bf 789}, 13 (2014)
  [arXiv:1402.2301 [astro-ph.CO]].

\bibitem{Boyarsky:2014jta}
  A.~Boyarsky, O.~Ruchayskiy, D.~Iakubovskyi and J.~Franse,
  Phys.\ Rev.\ Lett.\  {\bf 113}, 251301 (2014)
  [arXiv:1402.4119 [astro-ph.CO]].

\bibitem{Jeltema:2014qfa}
  T.~E.~Jeltema and S.~Profumo,
  arXiv:1408.1699 [astro-ph.HE].

\bibitem{constraint} A. Boyarsky {\it et al.}, Phys. Rev. Lett. {\bf
  102}, 201304 (2009).

\bibitem{model1} A. Kusenko, Phys. Rev. Lett {\bf 97}, 241301 (2006).

\bibitem{model2} F. Bezrukov, H. Hettmansperger and M. Lindner, Phys. Rev. D {\bf 81}, 085032 (2010).

\bibitem{Merle:2013gea}
  A.~Merle,
  Int.\ J.\ Mod.\ Phys.\ D {\bf 22}, 1330020 (2013)
  [arXiv:1302.2625 [hep-ph]].

\bibitem{pulsar} A. Kusenko and G. Segre, Phys. Lett. B {\bf 396}, 197
  (1997).

\bibitem{Hidaka1} J. Hidaka and G. M. Fuller, Phys. Rev. D {\bf 74},
  125015 (2006).
\bibitem{Hidaka2} J. Hidaka and G. M. Fuller, Phys. Rev. D {\bf 76},
  083516 (2007).

\bibitem{Bethe:1990mw}
  H.~A.~Bethe,
  Rev.\ Mod.\ Phys.\  {\bf 62}, 801 (1990).

\bibitem{formalism} G. Sigl and G. Raffelt, Nucl. Phys. B {\bf 406},
  423 (1993).

\bibitem{formalism1} G. Raffelt and G. Sigl, Astropart. Phys. {\bf 1}, 165
               (1993).

\bibitem{loss1} K. Kainulainen, J. Maalampi and J. T. Peltoniemi,
Nucl. Phys. B {\bf 358}, 435 (1991).

\bibitem{loss2} X. Shi and G. Sigl, Phys. Lett. B {\bf 323}, 360
               (1994).

\bibitem{loss3} K. Abazajian, G. M. Fuller and M. Patel, Phys. Rev. D {\bf 64}, 023501 (2001).

\bibitem{RZ} G. G. Raffelt and S. Zhou, Phys. Rev. D {\bf 83}, 093014
(2011).

\bibitem{CZ} B. Cao and S. Zhou, in preparation.
\end{thebibliography}
\end{document}